\documentclass{article}

\usepackage[preprint]{neurips_2026}

\usepackage[utf8]{inputenc} 
\usepackage[T1]{fontenc}    
\usepackage{hyperref}       
\usepackage{url}            
\usepackage{booktabs}       
\usepackage{amsfonts}       
\usepackage{nicefrac}       
\usepackage{microtype}      
\usepackage{xcolor}         
\usepackage{amsmath} 
\usepackage{amssymb}
\usepackage{graphicx}

\usepackage[acronym]{glossaries}
\makeglossaries

\newacronym{dit}{DiT}{Diffusion Transformer}
\newacronym{asamt}{aSAMt}{atomistic Structural Autoencoder Model temperature-conditioned}
\newacronym{ffn}{FFN}{Feed Forward Network}
\newacronym{lddt}{LDDT}{Local Distance Difference Test}
\newacronym{igso3}{IGSO(3)}{Isotropic Gaussian on SO(3)}
\newacronym{rope}{RoPE}{Rotary Position Encoding}
\newacronym{rg}{$R_g$}{Radius of Gyration}
\newacronym{rmsf}{RMSF}{Root Mean Square Fluctuation}
\newacronym{pdb}{PDB}{Protein Data Bank}

\newcommand{\gl}[1]{\gls{#1}}

\graphicspath{ {./figs/} }  
\title{Polyformer: a generative framework for thermodynamic modeling of polymeric molecules}

\author{
  Alessio Valentini$^1$\\
  \texttt{alessio@psidagger.ai} \\
  \And
  David Pekker$^{1,2}$\\
  \texttt{pekkerd@pitt.edu} \\
  \And
  Chungwen Liang$^{1,2}$ \\
  \texttt{chungwen.liang@psidagger.ai} \\
  \And
  Todd Martinez$^{1,3,4}$ \\
  \texttt{Todd.Martinez@stanford.edu} \\
  \And
  Swagatam Mukhopadhyay$^{1}$\\
  \texttt{swag@psidagger.ai} \\
  \And
  \\
  \vspace{-1.cm}\\
  $^1$ PsiDagger, San Diego, CA\\
  $^2$ Department of Physics and Astronomy, University of Pittsburgh, Pittsburgh, PA \\
  $^3$ Department of Chemistry and The PULSE Institute, Stanford University, Stanford, CA \\
  $^4$ SLAC National Accelerator Laboratory, Menlo Park, CA
  }

\begin{document}

\maketitle

\begin{abstract}
The classic paradigm of structural biology is that the sequence of a biomolecule (protein, nucleic acid, lipid, etc) determines its conformation (shape) which determines its biological function. Protein folding programs like AlphaFold address this paradigm by predicting the single best conformation given a sequence that defines the molecule. However, biomolecules are not static structures, and their conformational ensemble determines their function. We present the Polyformer -- a generative framework for thermodynamic modeling of polymeric molecules. Given the sequence and temperature (or another thermodynamic variable), the Polyformer generates conformations faithful to the molecule's thermodynamic conformational ensemble. It is the first generative model that solves three problems simultaneously: how does a molecule fold, what is its conformational ensemble, and how does the conformational ensemble change as we change physical temperature. As a concrete test case, we apply Polyformer to protein domains with 50-111 residues and report good agreement of model predictions to Molecular Dynamics (MD) trajectories.
\end{abstract}

\section{Introduction}
Large molecules, like proteins, nucleic acids, lipids, and various other polymers are dynamic rather than rigid structures. The conformational ensembles of these molecules, that is the collections of 3D shapes that these molecules explore, are critical for determining their physical, chemical, and biological properties. The field of molecular biology is undergoing a paradigm shift from the sequence-structure-function paradigm, where a single folded conformation determines the molecule's function, to the sequence-conformational ensemble-function paradigm, where the ensemble of conformations determines the function~\cite{Nussinov2023}. Some examples of the relation between function and conformation ensembles include mixing-unmixing transition of polymers grafted on a solid surface~\cite{deGennes1980}, intrinsically disordered proteins that derive their function from their `fuzzy' structure~\cite{Wright2015}, and function driven by large conformational changes in proteins~\cite{Orellana2019}.

Since the 1970's physics-based Molecular Dynamics (MD) and Monte Carlo models, that use either first-principles quantum mechanics calculations or parametrized inter-atomic potentials, have been used to study molecular conformations. The main advantage of these models is their capacity to describe the ensemble of molecular conformations as a function of thermodynamic variables like temperature and ion concentrations. However, these models are computationally very expensive, they can get stuck in local minima, and their ability to reproduce inter-atomic interactions at multiple relevant length-scales is inconsistent. As a result of these drawbacks the physics-based models have been unable to predict the shape of large molecules like proteins. Over the past half-decade, machine-learning approaches have revolutionized our understanding of molecular conformations. Specifically, programs like AlphaFold~\cite{Jumper2021} and RoseTTAFold~\cite{Baek2021} have made significant progress in solving the protein folding problem by generating a single best conformation given a protein sequence. More recently, it was realized that flow matching models could be used to generate multiple conformations for a given protein sequence and thus sample conformational ensembles~\cite{jing2024}. 

In this manuscript, we address the question whether foundation models can be endowed with the desirable properties of physics-based models for the task of thermodynamic conformational ensemble sampling, including the dependence of the ensemble on thermodynamic variables. We present the Polyformer, a generative machine-learning framework for sampling thermodynamic ensembles of conformations of polymeric molecules. At inference, our  model only needs the sequence and the temperature (or other thermodynamic observable that the model is trained on). Polyformer has a relatively simple architecture that was inspired by Simplefold~\cite{Wang2025}, which showed that one can avoid the PairFormer architecture of AlphaFold-2~\cite{Jumper2021} and RoseTTAFold~\cite{Baek2021}. Specifically, we use \gl{dit} in which the translations of the monomers is embedded using 3D Fourier embedding with learnable wave-vectors that capture multi-resolution and scale, while the rotations are embedded using Wigner-D matrices. We find that a custom temperature gating that modulates the input to attention and \gl{ffn} is critical for learning the temperature dependence, in contrast with time gating that acts on the output of attention and \gl{ffn} layer. We also find that a learnable chain polymer prior to bias attention that captures the polymer persistence length and its dependence on temperature is helpful, as is a loss function that includes the multi-scale \gl{lddt} loss for temperature ensembles. While we believe that the Polyformer architecture will work on different classes of molecules, in order to demonstrate its functionality we apply it to the problem of thermodynamic sampling of short proteins and peptides conformations. We use the ESM-2 model~\cite{lin2023evolutionary} to provide the initial embedding of the protein sequence. As we are most interested in flexible molecules, we train the Polyformer on a subset of the mdCATH dataset~\cite{Mirarchi2024} consisting of domains shorter than 111 amino acids.  We observe good agreement between the conformation ensembles, across temperatures, sampled by Polyformer and the mdCATH dataset. 

\section{Related Work}
Boltzmann sampling of molecular conformations using a machine learning framework originated with the Boltzmann generators~\cite{Noe2019}. It was soon realized that diffusion models could significantly improve performance and allow conformation sampling of large molecules~\cite{Wang2022}, thus resulting in machine-learned thermodynamic models of molecules. These initial efforts learned the thermodynamics of one molecule at a time, and thus while they were able to address the `jiggles and wiggles'~\cite{feynman2011feynman} as well as denaturation tasks, they did not aim to address the folding task. Many protein folding models have adopted diffusion~\cite{Abramson2024, Wohlwend2024} or flow matching~\cite{Wang2025} for generating conformations. While the underlying technology built into these models should be capable of sampling Boltzmann distribution of conformations, these models are typically not trained on Boltzmann ensembles due to lack of data.

There are two past thermodynamic models for large molecules. The \gl{asamt} model~\cite{Janson2025} works by taking an initial (low temperature) conformation of a protein and using it to generate conformational ensembles at different temperatures. The framework was trained on the mdCATH dataset~\cite{Mirarchi2024} that contains molecular dynamics trajectories for protein domains at different temperatures. The \gl{asamt} model was observed to correctly predict temperature-dependent changes in the conformational ensembles. The main drawback of the \gl{asamt} model, which the authors acknowledge in their paper, is that it relies on having a good initial conformation. That is the \gl{asamt} model learns the the `jiggles and wiggles' and how to denature the protein but not how to fold it. The other framework is RNAanneal~\cite{Herron2026}. RNAanneal uses classical secondary structure, tertiary structure, and molecular dynamics tools to propose 3D structure of RNA molecules. The Molecular dynamics data is fed to a machine learning thermodynamic model that estimates the free energy of the various 3D structures and thus assigns them a weight in the ensemble as a function of temperature. RNAanneal was observed to be successful at recovering experimentally observed conformations of 16 riboswitch RNAs. While RNAanneal is capable of performing all three tasks, including folding, it is significantly more computationally complex to run inference with RNAanneal than with generative models like \gl{asamt} and Polyformer.

\section{Methods}

\begin{figure}
    \centering
    \includegraphics[width=\linewidth]{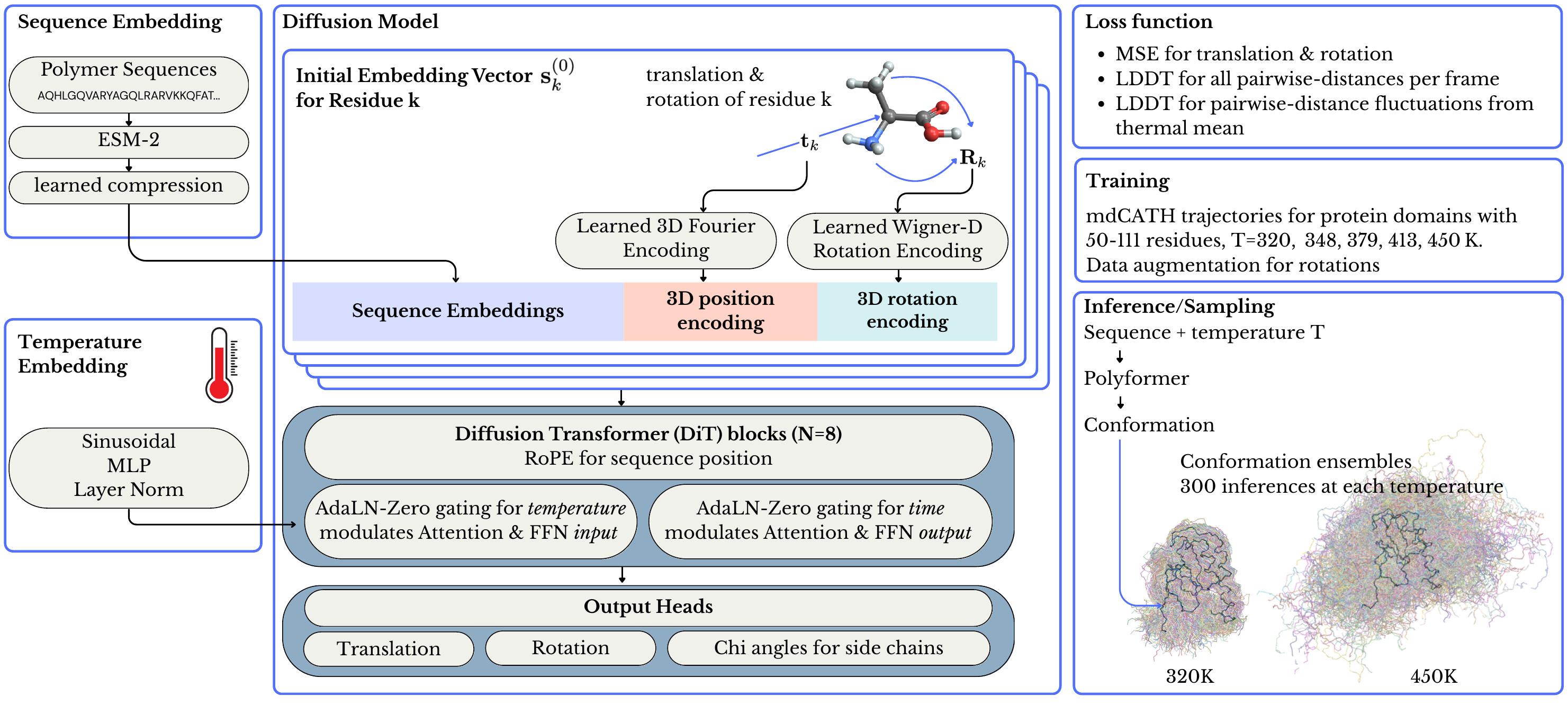}
    \caption{Polyformer is a \gl{dit} conditioned on molecular sequence and temperature that generates conformation ensembles of protein backbones in a temperature-dependent way. }
    \label{fig:high-level}
\end{figure}

Polyformer  architecture is schematically depicted in Fig.~\ref{fig:high-level}. 

\subsection{Input Encoding}
\label{sec:encoding}
Each residue $k$ is encoded into an initial embedding vector $\mathbf{s}_k^{(0)}$ by concatenating three feature blocks:
\begin{equation}
\mathbf{s}_k^{(0)} = \bigl[\,
  \text{RMSNorm}\left(\mathbf{W}_e\, \mathbf{e}_k\right)
  \;;\;
  \Phi(\mathbf{t}_k)
  \;;\;
  \Psi(\mathbf{R}_k)
\,\bigr].
\label{eq:node-encoder}
\end{equation}

\paragraph{ESM-2 feature encoding block}
We take the ESM-2 embedding vector $\mathbf{e}_k \in \mathbb{R}^{1280}$ and apply a learned linear projection $\mathbf{W}_e \in \mathbb{R}^{d_{\mathrm{esm}} \times 1280}$ followed by RMSNorm to bring the output to unit RMS scale. We choose $d_{\mathrm{esm}}=158$ so that ESM-2 features have roughly the same dimension as the geometry features. 

\paragraph{3D Fourier translation encoding block}
We encode residue translations, $\mathbf{t}_k \in \mathbb{R}^3$, using $M = 64$ learned 3D reciprocal vectors $\{\mathbf{k}_m \in \mathbb{R}^3\}_{m=1}^{M}$:
\begin{equation}
\Phi(t_k) = \bigl[\,
  \sin(\mathbf{k}_1 \cdot \mathbf{t}_k),\;
  \cos(\mathbf{k}_1 \cdot \mathbf{t}_k),\;
  \ldots,\;
  \sin(\mathbf{k}_M \cdot \mathbf{t}_k),\;
  \cos(\mathbf{k}_M \cdot \mathbf{t}_k)
\,\bigr] \in \mathbb{R}^{2M}.
\label{eq:fourier-3d}
\end{equation}
The $\mathbf{k}_m$'s, which are learnable, resolve spatial periods from ${\sim}126$\AA{} down to ${\sim}0.5$\AA{} at initialization. When two such encodings participate in a query-key dot product, the result contains terms of the form $\cos(\mathbf{k}_m \cdot (\mathbf{t}_i - \mathbf{t}_j))$ --- functions of the true 3D relative displacement between residues $i$ and $j$.  Making these wave vectors learnable parameters allows us to model structurally meaningful directions and Fourier components (e.g., helix axes, sheet normals, C$\alpha$--C$\alpha$ bond vectors).  

\paragraph{Wigner D-matrix rotation encoding block}
Resiudue rotations, $R_k \in \mathbb{R}^3$, are encoded via Wigner D-matrix features up to angular momentum order
$\ell_{\max} = 2$.  
The total rotation encoding is $\Psi(R_k) \in \mathbb{R}^{34}$ consisting of 9 features for $\ell = 1$ and 25 for $\ell = 2$. The Wigner D-matrices are the natural basis functions on $\mathrm{SO}(3)$
(Peter--Weyl theorem), and their bilinear products capture
relative rotations naturally.

\subsection{Forward Diffusion Process}
\label{sec:forward}

The forward process independently corrupts translations and rotations at
continuous time $t \in [0, 1]$, where $t = 0$ is clean and $t = 1$ is
maximally noised. Translations are measured in \AA{}  units and we use a noise schedule $\beta(t)$ that increases linearly from $\beta_{\min} = 0.1$ at $t=0$ to $\beta_{\max} = 20.0$ at $t=1$. Rotations are perturbed by the \gl{igso3} distribution~\cite{leach2022}, parameterized by a logarithmic sigmoid noise schedule with $\sigma_{\min} = 0.1$ at $t=0$ and $\sigma_{\max} = 1.5$ at $t=1$. Details of forward diffusion process appear in Appendix~\ref{app:forward}.

\subsection{Diffusion Model Architecture and Conditioning}
\label{sec:conditioning}

\paragraph{Temperature and diffusion timestep encoding}
The diffusion timestep $t \in [0, 1]$ and normalized temperature
$T_{\text{norm}} \in [0, 1]$ are embedded via independent sinusoidal
encoding $\to$ MLP pathways, each followed by non-affine LayerNorm. See Appendix~\ref{app:cond} for details.

\paragraph{\gl{dit} blocks with factored AdaLN}
The model trunk consists of $N=8$ \gl{dit} blocks. Each block contains a self-attention sublayer and a SwiGLU feed-forward sublayer, both modulated by timestep and temperature through an adaptive layer normalization scheme.  Each sublayer is preceded by parameter-free LayerNorm, denoted $\mathrm{LN}$, whose missing scale and bias are replaced by the adaptive modulation parameters described next. Inside the attention sublayer, queries and keys undergo per-head RMSNorm---which normalizes to unit RMS then applies a learnable per-channel scale---before \gl{rope} and the dot product. This decouples Q/K direction from input magnitude, letting the network learn attention scales independently of representation norm. We include a chain-proximity bias in the attention, which helps to inform the model of the polymer proximity length, see Appendix~\ref{app:chain-prox} for details.

\emph{Timestep modulation} produces 6$d$ parameters
$(\beta_a, \gamma_a, g_a^{(t)}, \beta_f, \gamma_f, g_f^{(t)})$
from $c_t$ via a zero-initialized linear projection, providing AdaLN-Zero~\cite{Peebles2023}
modulation (shift, scale, gate) for each sublayer.

\emph{Temperature modulation} produces $4d + 2$ parameters from $c_T$ via a
randomly initialized linear projection (std $= 0.02$): full AdaLN
$(\gamma_a^{(T)}, \beta_a^{(T)}, \gamma_f^{(T)}, \beta_f^{(T)})$
for each sublayer, plus two chain-prior scalars $(\tilde{a}, \tilde{\gamma})$.

One of the key design choices we made was that the two signals act at different structural points.  For the attention sublayer,
temperature modulates the sublayer \emph{input} while timestep gates the
sublayer \emph{output}:
\begin{align}
\hat{\mathbf{s}} &= \mathrm{LN}(\mathbf{s}) \odot (1 + \gamma_a) + \beta_a,
  \label{eq:adaln-t} \\
\tilde{\mathbf{s}} &= \hat{\mathbf{s}} \odot (1 + \gamma_a^{(T)}) + \beta_a^{(T)},
  \label{eq:adaln-T} \\
\mathbf{s} &\leftarrow \mathbf{s} + g_a^{(t)} \odot \mathrm{Attn}(\tilde{\mathbf{s}}),
  \label{eq:gate}
\end{align}
with an identical pattern for the \gl{ffn} sublayer.  The effective per-channel
modulation composes multiplicatively in scale and additively in shift:
$\text{scale}_{\mathrm{eff}} = (1 + \gamma_t)(1 + \gamma_T)$ and
$\text{shift}_{\mathrm{eff}} = \beta_t (1 + \gamma_T) + \beta_T$.
Because the timestep gate $g_a^{(t)}$ is zero-initialized, the block is the identity at initialization regardless of the temperature modulation, ensuring stable early training. The temperature projection weights are randomly initialized (std $= 0.02$), so that once the gates open, different temperatures immediately produce distinct modulations without a symmetry-breaking phase.

\subsection{Output Heads}
\label{sec:output}

The final layer applies non-affine LayerNorm to the trunk output, followed by
the same factored AdaLN modulation (timestep shift/scale, then temperature
shift/scale), and projects through three separate linear heads:

\begin{itemize}
\item \textbf{Translation head} (3D): zero-initialized, producing
  $\Delta t_k$ (residual update to the noisy translation).
\item \textbf{Rotation head} (3D): randomly initialized (std $= 0.02$),
  producing a rotation vector $\mathbf{v}_k$ (axis-angle).
\item \textbf{$\chi$ angle head} ($2K$-D, $K = 4$): producing $(\sin \chi, \cos \chi)$ pairs for up to 4 sidechain dihedral angles, depending on residue (e.g Alanine has no $\chi$'s while Arginine has four).
\end{itemize}

Clean frames are reconstructed as:
\begin{equation}
\hat{t}_0 = t_t + \Delta t_k, \qquad
\hat{R}_0 = \exp([\mathbf{v}_k]_\times) \cdot R_t,
\label{eq:reconstruction}
\end{equation}
where $\exp([\cdot]_\times)$ denotes the Rodrigues formula applied to the
predicted rotation vector.

\subsection{Loss Function}
\label{sec:loss}

The total loss is a weighted sum of five terms:
\begin{equation}
\mathcal{L} =
  \mathcal{L}_{\mathrm{trans}} + \mathcal{L}_{\mathrm{rot}}
  + \lambda_\chi \mathcal{L}_\chi
  + \lambda_{\mathrm{lddt}} \mathcal{L}_{\mathrm{lddt}}
  + \lambda_{\mathrm{ens}} \mathcal{L}_{\mathrm{ens}}.
\label{eq:total-loss}
\end{equation}
The first four terms correspond to translation, rotation, $\chi$-angle, and standard \gl{lddt} loss. These are conventional loss terms and are described in Appendix~\ref{app:loss}. 

To provide an explicit temperature-dependent supervisory signal, we introduce an ensemble \gl{lddt} loss, $\mathcal{L}_{\mathrm{ens}}$, that compares predicted pairwise C$\alpha$ distances at a specific temperature against
pre-computed ensemble mean distances $\mu_{kl}^{(T)}$ at that temperature $T$.  Given a set of wider thresholds
$\mathcal{T}_{\mathrm{ens}} = \{1.0, 2.0, 4.0, 8.0, 16.0, 32.0\}$\AA{}:
\begin{equation}
\mathcal{L}_{\mathrm{ens}} = 1 -
  \frac{1}{|\mathcal{P}|}
  \sum_{(k,l) \in \mathcal{P}}
  \frac{1}{|\mathcal{T}_{\mathrm{ens}}|}
  \sum_{\tau \in \mathcal{T}_{\mathrm{ens}}}
  \sigma\!\Bigl(
    \frac{\alpha}{\tau}
    \bigl(\tau - |d_{kl}^{\mathrm{pred}}{}^{(T)} - \mu_{kl}^{(T)}|\bigr)
  \Bigr),
\label{eq:loss-ens}
\end{equation}
where $\alpha = 10$ is the sigmoid slope and $\mathcal{P}$ is the set of valid
residue pairs.  The ensemble means $\mu_{kl}^{(T)}$ are computed offline from MD trajectories at each temperature.

The thresholds are deliberately much wider than those of the standard \gl{lddt} loss, reflecting the fact that temperature-induced distance
changes span a broader range than single-frame reconstruction errors.  Rigid
pairs---where $\mu_{kl}^{(T)}$ barely varies with $T$---pass all thresholds
trivially and contribute zero gradient.  The temperature signal therefore arises
exclusively from flexible pairs whose ensemble mean distances shift
meaningfully with $T$, giving the model a direct learning target for
temperature-dependent conformational changes.  The loss is bounded in $[0, 1]$
by construction, requiring no additional normalization or variance weighting.


\subsection{Model training}
We train with AdamW~\citep{loshchilov2019} (lr $= 2 \times         
10^{-4}$, weight decay $= 0.01$, gradient clipping $= 1.0$) for $10^6$ steps with cosine annealing after 2,000 warmup steps. Batch size is 48 on a single
NVIDIA RTX 3090 (24\,GB) with fp16 mixed precision.  Exponential moving
average (EMA) with decay $0.999$ is used for evaluation.  In the loss function we set $\lambda_{\chi}=0.1$, $\lambda_{\mathrm{lddt}}=1.0$, and $\lambda_{\mathrm{ens}}=1.0$. Model training required 2 days.
\paragraph{Data}
We train on mdCATH molecular dynamics ensembles (320--450\,K, 5 temperatures) filtered to $\leq 111$ residues. The training set consists of 2,142 domains, and the testing set of 103 domains. Train/validation splits use CATH superfamily grouping to prevent structural similarity leakage. ESM-2 embeddings are precomputed and cached. SE(3) augmentation is applied per step, and translations are centered to zero mean. 

\subsection{Inference/conformation sampling (DDIM)}
\label{sec:inference}
Conformations are generated by reverse diffusion from $t = 1$ to $t \approx 0$ over $S = 100$ uniformly spaced steps using DDIM with $\eta = 1$ (fully stochastic, equivalent to DDPM). For each domain and temperature we generate 300 conformations. See Appendix~\ref{app:inference} for details of the reverse diffusion.

\section{Results}
\begin{figure}
    \centering
    \includegraphics[width=0.7\linewidth]{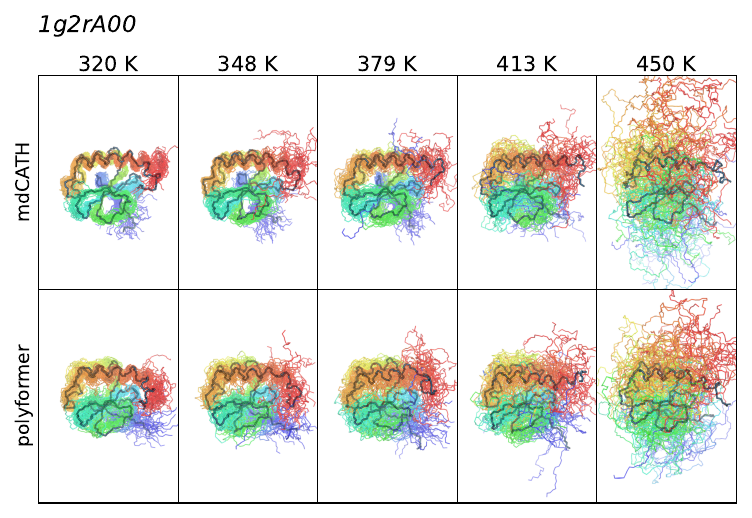}
    \caption{Comparison of Molecular Dynamics and Polyformer sampled conformation ensembles for domain 1g2rA00. Best matching pair of conformations is highlighted. The ensembles of conformations sampled by Polyformer are quite similar to Molecular Dynamics across all five temperatures. 
    }
    \label{fig:potato}
\end{figure}

In order to get a qualitative understanding for how polyformer performs, we compare mdCATH and polyformer conformations for domain lg2Ar00, see Fig.~\ref{fig:potato}. For each of the five temperatures in the mdCATH dataset, we plot 100 conformations from the mdCATH dataset and by running polyformer.  For each temperature, we highlight the pair of conformations that are most similar in mdCATH and polyformer samples as measured by RMSD (highlighted in the Figure). First, we observe that lg2Ar00 has a lot of fluctuations and would be poorly described by a single conformation even at the lowest temperatures. Second, we observe that the highlighted conformations are very similar for mdCATH and polyformer at all temperatures indicating that Polyformer is capable of generating structures that match molecular dynamics. Third, we observe that as the temperature increases from 320K to 450K, the domain becomes denatured and the ensemble of conformations becomes more and more disorder and the \gl{rg} grows. This increase in disorder and \gl{rg} tracks well between mdCATH and Polyformer. These behaviors are typical of polyformer performance, see Appendix~\ref{app:more-data} for another typical domain, 3g0vA00.

\begin{figure}
    \centering
    \includegraphics[width=\linewidth]{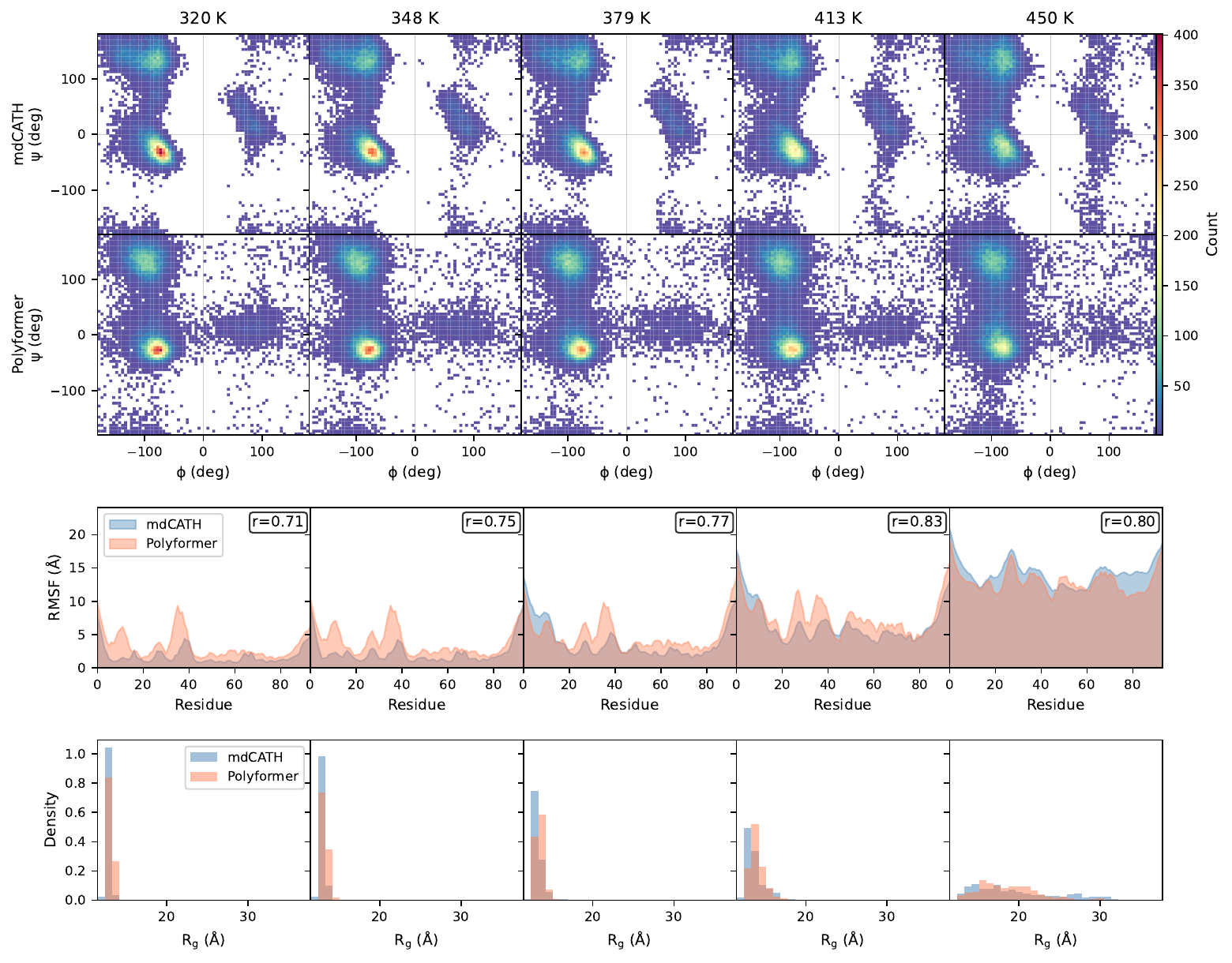}
    \caption{Comparison of Molecular Dynamics and Polyformer  sampled conformation ensemble for domain 1g2rA00. (a) Ramachandran plots comparing the distribution of $\psi$ and $\phi$ dihedral angles. (b) RMSF plots comparing fluctuations along the back bone. (c) Plots comparing distributions of \gl{rg}. Across all three metrics, polyformer matches molecular dynamics pretty well.}
    \label{fig:single_domain_dist}
\end{figure}

\begin{figure}
    \centering
    \includegraphics[width=0.5\linewidth]{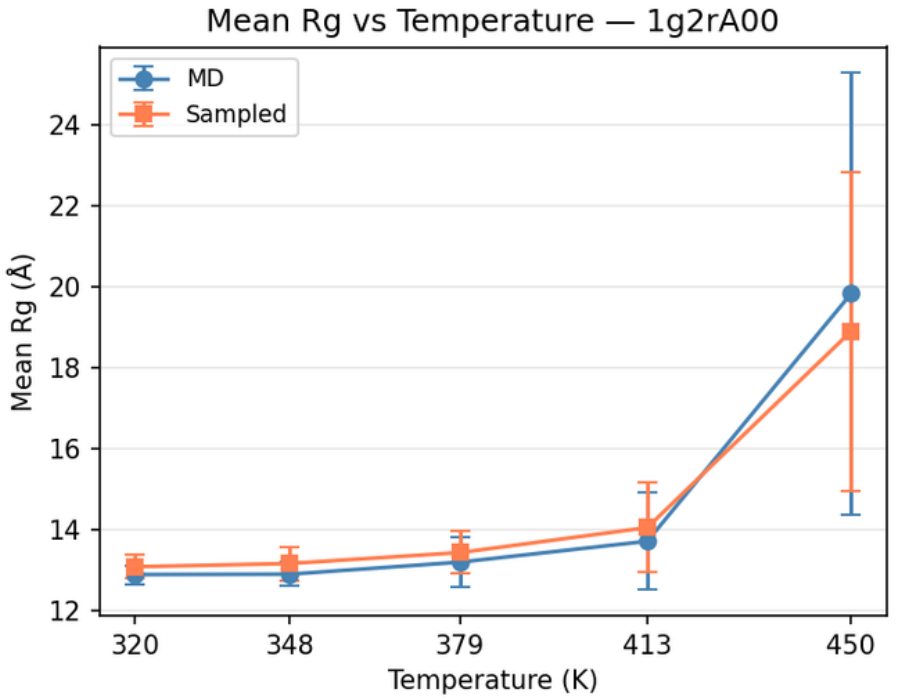}
    \caption{Comparison of mean (solid points) and standard deviation (error bars) of \gl{rg} obtained by mdCATH and Polyformer ensemble sampling of domain lg2A00, showing good agreement.}
    \label{fig:single_domain_rg}
\end{figure}

Next, we move on to a quantitative comparison of the conformation ensembles. We begin by comparing the backbone dihedral angles ($\phi, \psi$) in Ramachandran plots, see Fig.~\ref{fig:single_domain_dist}a. For each domain and temperature, angles were pooled across all structures in the ensemble and binned into two-dimensional histograms with $5^\circ$ resolution (72 bins per axis, spanning $-180^\circ$ to $180^\circ$). Bins with fewer than one count are masked. All panels share a common color scale, enabling direct visual comparison of secondary-structure populations between the mdCATH reference and Polyformer-generated ensembles. We observe that the weight on the main $\alpha$ and $\beta$ peaks decrease with temperature, with weight on the $\beta$ peak decreasing faster. These trends match well between mdCATH and Polyformer conformation ensembles. However, the shape of the main peaks as well as satellite peaks appears to be somewhat different indicating imperfections in either the Polyformer generated structures or reconstructed dihedral angles. Next we look at \gl{rmsf} along the chain of the protein, see Fig.~\ref{fig:single_domain_dist}b. First, we observe that the size of fluctuations increases with temperature, whatever structure this domain had at 320K becomes denatured at 450K. Second, we observe that the \gl{rmsf} traces obtained from mkdCATH and Polyformer match well across all five temperatures. Finally, we look at the distributions of \gl{rg}, see Fig.~\ref{fig:single_domain_dist}c. First, we observe that the mean and the width of the distribution increases with temperature indicating that the domain is becoming denatured. Second, we observe pretty good correspondence between mdCATH and  Polyformer sampled \gl{rg} distribution. Finally, we compare the mean and the standard deviation of the \gl{rg} distributions, as a function of temperature, in Fig.~\ref{fig:single_domain_rg}, and again find good agreement. Taken together, these observations indicate that the Polyformer is doing a reasonably good job of sampling the conformation ensemble. See Appendix~\ref{app:more-data} for another typical domain, 3g0vA00.

\begin{figure}
    \centering
    \includegraphics[width=\linewidth]{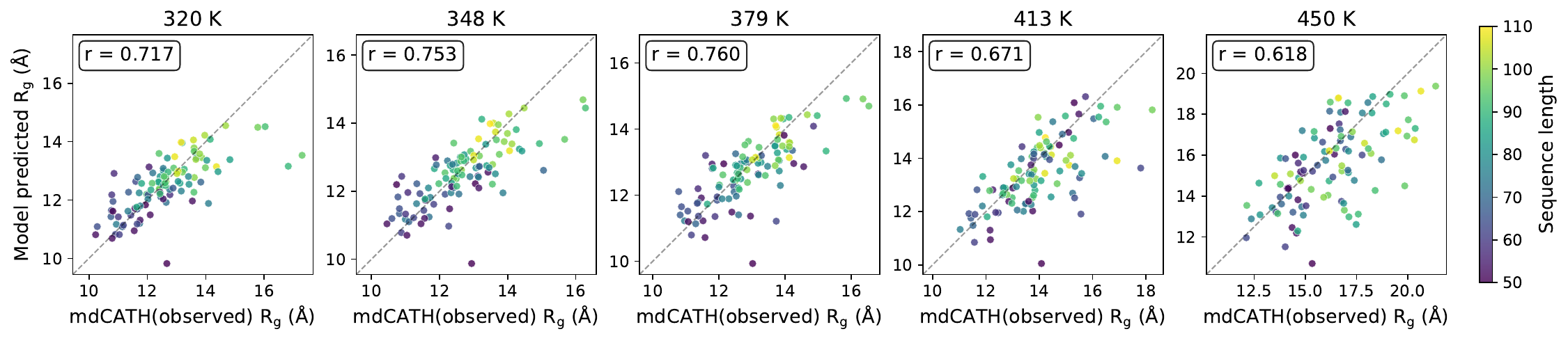}
    \caption{Comparison of mean \gl{rg} obtained by mdCATH and Polformer ensemble sampling across all domains in the testing set and all temperatures in mdCATH dataset. The plots show good, but imperfect correlation. }
    \label{fig:all_domains_rg}
\end{figure}

Finally, we look at the performance of the Polyformer across the testing set. Here, we compare the mean value of \gl{rg} obtained by mdCATH and Polformer sampling, see Fig.~\ref{fig:all_domains_rg}. We observe strong correlation at lower temperatures that becomes not quite as strong at high temperatures.

\section{Discussion}

Our goal in constructing the Polyformer was to make it a general framework for learning ensembles of polymer conformations, and their dependence on thermodynamic variables, from datasets of MD trajectories. To that end, we have made several deliberate choices in the the model architecture.
First, we chose a flexible approach to treating the monomers that make up the polymer.  For each monomer we specify its position and a set of internal degrees of freedom. For the protein case, we specified the position of the monomer by specifying the position of the $C_\alpha$ atom and we used internal degrees of freedom to specify the 3D rotational orientation of the monomer and the $\chi$-angles of the side chain. Polyformer can be extended to contain additional/different internal degrees of freedom like relative position of all atoms that make up the monomer.
Second, we incorporated a polymer chain prior as a physical bias to attention mechanism (with learnable contribution and decay exponent of monomer separation along the polymer, see Appendix~\ref{app:chain-prox}). 
Third, we abandoned the Pairformer architecture in favor of a much richer learnable Fourier Embeddings in 3D space~\cite{li2021learnable}. This significantly reduces our model size and the complexity. 
The learnable Fourier modes in our model indeed learn multiple Fourier modes physically meaningful for polymer length-scales, see Appendix~\ref{app:fourier}. 
Fourth, we implemented temperature gating in \gl{dit} in a physically meaningful way. Temperature should shape how the model attends to the monomers and should be independent of how diffusion time modulates the influence of the attention block. We achieved this by temperature-gating the input in attention block, whereas time-gating applied to the output as is conventional in \gl{dit}. 
Fifth choice, though conventional, is worth discussing---LDDT losses captures the multi-scale conformational properties of polymers by comparing distances between pairs of monomers. We apply LDDT loss in two distinct paths. The first path is the conventional LDDT loss, that penalizes the difference between two conformations. The second path, which we introduce, penalizes the difference between a pairwise distance at a temperature and the empirical ensemble mean of that distance at that temperature over all the MD trajectories. This loss also acts at multiple scales, but is more relaxed in overall scale compared to frame LDDT loss, capturing the thermal ensemble flexibility per monomer pair. The two losses compete with each other---the first one demands frame level fidelity, while the second one demands fidelity of thermal averages, both at multiple and distinct multi-scales. Thereby, the ensemble loss provides a first-moment supervision signal for temperature dependence, complementing the distributional learning that the frame-level losses provide.

Polyformer is a Foundation Models of polymeric molecules that is designed to addressed three different learning tasks. These are: (a) folding of the sequence to a single conformation, (b) sampling the conformation ensemble, and (c) describing how the conformation ensemble depends on temperature. AlphaFold and RoseTTAFold-family of models handle task (a), AlphaFlow-family of models handle tasks (a) and (b), Boltzmann generators, previous thermodynamic models, and aSAMt handle tasks (b) and (c). The only other work that we are aware of that handles all three tasks is RNAnneal~\cite{Herron2026}. However, unlike Polyformer, RNAnneal is not a fully  generative model---it uses physics-based models to propose conformations and a machine learning thermodynamic model to score the conformations. 

When comparing the performance Polyformer to other models, it is important to recognize that: (a) Polyformer has a harder learning objective as it was designed to achieve all three tasks and (b) Polyformer was trained on a dataset that contained 3D conformations for only 2,142 protein sequences as compared to roughly 170,000 sequences with 3D structure from the \gl{pdb} that were used for training by AlphaFold and RoseTTAFold. While a conformational ensembles are a lot more information rich as compared to just a single conformation from the \gl{pdb}, we found it surprising that Polyformer was able to learn how to fold protein domains with so little 3D data. Polyformer does use the ESM-2 protein language model embeddings, which do not carry any specific 3D conformation data. We did perform an ablation study to test the influence of ESM-2 encoding on Polyformer's performance and found that there is only a moderate decrease in performance, see Appendix~\ref{app:no-ESM2}. Polyformer demonstrates that protein structure can indeed be learned from a rather moderate dataset.

Our choice of proteins as the first target system for Polyformer was dictated by the availability of the mdCATH dataset~\cite{Mirarchi2024}. Reference MD data, that samples not just a single conformation per sequence, but the whole thermodynamic ensemble at different temperatures is crucial for training Polyformer. Indeed, this is well demonstrated by comparing Polyfromer to SimpleFold which was trained on \gl{pdb} data. While SimpleFold has the cpabaility to predict different protein conformations, in practice it fails to predict a thermodynamic distribution and instead tends to predict just a single dominant conformation, see Appendix~\ref{app:simplefold}.

The mdCATH dataset in its present form, while invaluable, does have limitations that limit the performance of PolyFormer. First, the dataset includes only 5,398 protein domains (sequences), of which we used a subset of 2,142 for training and 103 for testing. Second, the mdCATH dataset is biased towards small, ordered domains. Limitations of state-of-the-art force field molecular dynamics dictate that domain size cannot be larger than 500 residues. Further, the \gl{pdb}, where the mdCATH domains originate, itself is biased towards well-structured protein and away from disordered proteins/peptides. It would be extremely valuable to generate additional MD data on disordered domains to supplement mdCATH. Third, the MD trajectories sampled by mdCATH are not fully ergodic---they do not sample the whole Boltzmann ensembles for protein domains, especially at lower temperatures. Ergodic sampling using MD is a challenging problem because trajectories tend to get stuck in minima of the free energy landscape. There are various important sampling strategies, like replica exchange MD~\cite{sugita1999replica}, umbrella sampling~\cite{torrie1977nonphysical}, matrix product state exploration~\cite{han2018unsupervised}, that attempt to improve ergodicity. Perhaps in the future, more ergodic MD trajectories could be generated using one or more of these strategies. Finally, mdCATH relies on CHARMM22* force field. While this is the state-of-the-art Force Fields for proteins, it does tend to predict denaturation temperatures that are too high~\cite{huang2017charmm36m}. We expect that performance of such generative thermodynamic conformation ensemble models will continue to improve as Force Fields become more accurate and provide us with more accurate training data over a wider range of sequences.

What is Polyformer actually learning? Previous work on thermodynamic models show that predicting thermal ensembles is intimately tied to predicting the Free Energy landscape, and its dependence on temperature~\cite{Wang2022,Herron2026}. Polyformer does raise the tantalizing possibility of being able to predict free energy from sequence alone---and perhaps it is not more data on frozen structures but higher quality datasets on the dynamics of conformations that is needed to ground Foundation Models of polymers to Physics. We suspect that Polyformer itself can be used as a tool to create better importance sampling for high-quality MD trajectories, leading to a flywheel effect in active learning of free energy of polymers.

\section{Conclusions}
We have demonstrated that it is possible to construct a single, generative, thermodynamic conformation ensemble sampling model of large molecules that learns how to fold these molecules, how these molecules thermally fluctuate, and how these molecules denature with increasing temperature. While the focus of the current paper is on protein domains, we believe that the same methodology can be applied to any class (or mixed classes) of molecules. We envision that Polyformer can be trained on several thermodynamic variables and thus it could be used to model how conformational ensembles change with the environment of the molecule, e.g. change of conformation under solvent conditions. Polyformer is the latest member of the new class of models that sample conformational ensembles, as opposed to trying to predict one best conformation, and thus can help to discover mechanism of function of peptides and other polymeric molecules.

\bibliographystyle{unsrt}
\bibliography{refs}

\newpage

{\Large Appendix}
\appendix
\section{Forward diffusion}
\label{app:forward}
\paragraph{Translation noise (VP-SDE, linear $\beta$).}
Translations are first scaled by a coordinate scaling factor
$s = 0.1$ to bring \AA ngstr\"om coordinates into unit range.  The
instantaneous noise rate is $\beta(t) = \beta_{\min} + t(\beta_{\max} -
\beta_{\min})$ with $\beta_{\min} = 0.1$, $\beta_{\max} = 20.0$.  The
cumulative noise is:
\begin{align}
\bar{\beta}(t) &= t \,\beta_{\min} + \tfrac{1}{2}\, t^2\, (\beta_{\max} - \beta_{\min}), \\
\bar{\alpha}(t) &= \exp \bigl(-\bar{\beta}(t)\bigr).
\end{align}
The noisy translation at time $t$ is sampled as:
\begin{equation}
x_t = \sqrt{\bar{\alpha}(t)} \cdot x_0 + \sqrt{1 - \bar{\alpha}(t)} \cdot \epsilon, \qquad
\epsilon \sim \mathcal{N}(0, I),
\label{eq:forward-trans}
\end{equation}
where $x_0 = s \cdot t_k$ is the centered, scaled clean translation.

\paragraph{Rotation noise \gl{igso3}.}
Rotations are perturbed by the \gl{igso3}, parameterized by a
logarithmic sigma schedule:
\begin{align}
\sigma(t) &= \log\!\bigl(t \cdot e^{\sigma_{\max}} + (1 - t) \cdot e^{\sigma_{\min}}\bigr),
  \label{eq:sigma-sched} \\
\varepsilon(t) &= \sigma(t)^2 / 2,
  \label{eq:igso3-eps}
\end{align}
with $\sigma_{\min} = 0.1$ and $\sigma_{\max} = 1.5$.  The \gl{igso3} distribution
has density over rotation angle $\omega \in [0, \pi]$:
\begin{equation}
p(\omega \mid \varepsilon) \propto
  (1 - \cos \omega)
  \sum_{\ell=0}^{L}
  (2\ell + 1)\, e^{-\ell(\ell+1)\varepsilon}\,
  \frac{\sin\bigl((\ell + \tfrac{1}{2})\omega\bigr)}{\sin(\omega / 2)},
\label{eq:igso3-pdf}
\end{equation}
truncated at $L = 1000$ terms.  A rotation angle $\omega$ is sampled via inverse
CDF on a 1024-point grid, then combined with a uniformly random axis on $S^2$
via Rodrigues' formula to yield the perturbation
$\Delta R \sim \mathrm{IGSO}(3)(\varepsilon)$.

\paragraph{Precomputed \gl{igso3} cache.}
Na\"ively, each call to sample from \gl{igso3} or compute the rotation SNR
(Eq.~\ref{eq:snr-rot}) requires evaluating the $L = 1000$-term Fourier series. We eliminate this cost by precomputing a lookup table at
startup: 512 CDF tables at log-spaced $\varepsilon$ values covering the
schedule's range, plus the corresponding $\mathbb{E}[\omega^2]$ values.
At runtime, rotation sampling reduces to a nearest-neighbor CDF lookup
followed by \texttt{searchsorted}, and $\mathbb{E}[\omega^2]$ is obtained
by linear interpolation in $\log\varepsilon$ space.  Log spacing provides
uniform relative precision---the CDF shape changes much
faster at small $\varepsilon$ (concentrated near 0) than at large
$\varepsilon$ (nearly uniform on SO(3)).  With 512 grid points the
quantization error is $< 0.2\%$ relative.

The noisy rotation is:
\begin{equation}
R_t = \Delta R \cdot R_0.
\label{eq:forward-rot}
\end{equation}

\section{Dual conditioning pathway for temperature and diffusion timestep}
\label{app:cond}

\paragraph{Temperature normalization.}
Physical temperature $T_K$ (Kelvin) is canonically normalized as $T_{\mathrm{norm}} = (T_K - 300) / 200$, mapping the training range $320$--$450$\,K to $[0.1, 0.75]$ with room for extrapolation on both sides.

\paragraph{Embedding.}
The diffusion timestep $t \in [0, 1]$ and normalized temperature $T_{\text{norm}} \in [0, 1]$ are embedded via 
\begin{align}
c_t &= \mathrm{LN}\!\bigl(\mathrm{MLP}_t(\mathrm{SinEmb}(100 \cdot t))\bigr), \\
c_T &= \mathrm{LN}\!\bigl(\mathrm{MLP}_T(\mathrm{SinEmb}(100 \cdot T_{\mathrm{norm}}))\bigr),
\end{align}
where $\mathrm{SinEmb}(\cdot)$ is the standard sinusoidal positional encoding
with maximum period 1000, and each MLP is a two-layer network with SiLU
activation.  The factor of 100 spreads the $[0,1]$ input across the frequency
grid.  The non-affine LayerNorm (zero mean, unit variance, no learnable
scale/bias) normalizes each conditioning vector to a fixed norm, preventing one
pathway from dominating the other and forcing each MLP to encode information in
its \emph{direction}, not its magnitude.

\section{Chain-Proximity Bias (Polymer Prior in Attention)}
\label{app:chain-prox}

Temperature modulation additionally outputs two scalars
$(\tilde{a}, \tilde{\gamma})$ that parameterize a temperature-dependent polymer
chain prior added to the attention logits.  After softplus to ensure positivity:
\begin{equation}
a = \mathrm{softplus}(\tilde{a}), \qquad
\gamma = \mathrm{softplus}(\tilde{\gamma}),
\end{equation}
the chain-proximity bias matrix is:
\begin{equation}
C_{kl}(T) = \frac{a}{(|k - l| + 1)^{\gamma}},
\label{eq:chain-bias}
\end{equation}
where $k, l$ are residue indices.  This bias is added to the attention logits
before softmax, broadcast identically across all attention heads.

The power-law form $a / d^\gamma$ encodes the physical prior that nearby
residues along the polymer chain interact more strongly.  The exponent $\gamma$
controls the decay rate: $\gamma \to 0$ yields a uniform (no-prior) limit,
$\gamma \approx 1$ gives $1/d$ decay characteristic of polymer chains, and
$\gamma \to \infty$ produces a sharp nearest-neighbor cutoff.  Because both $a$
and $\gamma$ are predicted from the temperature conditioning vector $c_T$, the
model can learn temperature-appropriate structural priors: compact states
(low $T$) may favor stronger local bias, while flexible states (high $T$) may
reduce the magnitude and steepen or flatten the decay.

This bias is complemented by \gl{rope}, applied to
queries and keys after per-head RMSNorm.  \gl{rope} uses standard sinusoidal
frequencies $\theta_j = \text{base}^{-2j/d_h}$ with base $= 10000$, encoding
1D sequence separation (residue index) only.  Geometric (3D) information is
handled by the Fourier and Wigner input encodings and the chain-proximity bias,
not by \gl{rope}.

\section{Loss}
\label{app:loss}
\paragraph{Translation loss.}
Per-residue squared Euclidean distance in scaled coordinate space, weighted
inversely by the translation signal-to-noise ratio with min-SNR-$\gamma$
truncation~\cite{hang2023efficient}:
\begin{equation}
\mathcal{L}_{\mathrm{trans}} =
  \frac{1}{N}\sum_k
  \frac{\lambda_{\mathrm{rel}}}{\max\!\bigl(\mathrm{SNR}_{\mathrm{trans}}(t),\; \gamma\bigr)}
  \| \hat{t}_{0,k} - t_{0,k} \|^2,
\label{eq:loss-trans}
\end{equation}
where $\mathrm{SNR}_{\mathrm{trans}}(t) = \bar{\alpha}(t) / (1 - \bar{\alpha}(t))$
and $\gamma = 5$.  The clamping caps the per-timestep weight at $1/\gamma = 0.2$,
preventing extreme translation gradients near $t = 1$ where
$\mathrm{SNR}_{\mathrm{trans}} \to 0$.

\paragraph{Rotation loss.}
Squared chordal (Frobenius) distance on $\mathrm{SO}(3)$, also SNR-weighted:
\begin{equation}
\mathcal{L}_{\mathrm{rot}} =
  \frac{1}{N}\sum_k
  \frac{1}{\max\!\bigl(\mathrm{SNR}_{\mathrm{rot}}(t),\; \gamma\bigr)}
  \, d_F^2(\hat{R}_{0,k},\, R_{0,k}),
\label{eq:loss-rot}
\end{equation}
where $d_F^2(R, R') = 2(3 - \operatorname{tr}(R^\top R')) \in [0, 8]$.
This avoids the arccos singularities of the geodesic distance at $\omega = 0$
and $\omega = \pi$, providing smooth polynomial gradients everywhere.

The rotation SNR is defined analogously to translation:
\begin{equation}
\mathrm{SNR}_{\mathrm{rot}}(t) =
  \frac{\mathbb{E}_{\mathrm{unif}}[\omega^2] - \mathbb{E}_\varepsilon[\omega^2]}
       {\mathbb{E}_\varepsilon[\omega^2]},
\label{eq:snr-rot}
\end{equation}
where $\mathbb{E}_{\mathrm{unif}}[\omega^2] = \pi^2/3 + 2$ is the expected
squared rotation angle under the uniform distribution on $\mathrm{SO}(3)$, and
$\mathbb{E}_\varepsilon[\omega^2]$ is computed from the \gl{igso3} density at
the current $\varepsilon(t)$.

\paragraph{Chi angle loss.}
For sidechain dihedral angles predicted as $(\sin\chi, \cos\chi)$ pairs:
\begin{equation}
\mathcal{L}_\chi =
  \frac{1}{|\mathcal{V}|}\sum_{k,j \in \mathcal{V}}
  \bigl(1 - \cos(\hat{\chi}_{k,j} - \chi_{k,j})\bigr),
\label{eq:loss-chi}
\end{equation}
where $\mathcal{V}$ is the set of valid chi angles.  The $1 - \cos(\Delta)$
form naturally handles angular periodicity and ranges in $[0, 2]$.

\paragraph{Standard \gl{lddt} loss.}
A differentiable approximation to the \gl{lddt} score applied to predicted
pairwise C$\alpha$ distances, preventing structural fragmentation:
\begin{equation}
\mathcal{L}_{\mathrm{lddt}} = 1 -
  \frac{1}{|\mathcal{P}|}\sum_{(k,l) \in \mathcal{P}}
  \frac{1}{|\mathcal{T}|}
  \sum_{\tau \in \mathcal{T}}
  \sigma\!\Bigl(\frac{10}{\tau}\bigl(\tau - |d_{kl}^{\mathrm{pred}} - d_{kl}^{\mathrm{true}}|\bigr)\Bigr),
\label{eq:loss-lddt}
\end{equation}
where $\sigma$ is the sigmoid function, $\mathcal{P}$ is the set of valid
residue pairs, $\mathcal{T} = \{0.5, 1.0, 2.0, 4.0\}$\AA{}.

\section{Conformation generation}
\label{app:inference}
\paragraph{Initialization.}
Translations are drawn from the standard normal distribution: $x_1 \sim \mathcal{N}(0, I)$ (in {\AA} units).  Rotations are drawn
uniformly on SO(3) via \gl{igso3} with $\varepsilon = 100$ (effectively uniform).

\paragraph{Translation step (DDIM).}
At each step from $t_i$ to $t_{i+1}$ (with $t_{i+1} < t_i$), we use the DDIM
update~\cite{song2020denoising}:
\begin{align}
\hat{x}_0 &= x_t + \Delta t_{\mathrm{pred}}, \\
\hat{\epsilon} &= \frac{x_t - \sqrt{\bar{\alpha}_t}\, \hat{x}_0}{\sqrt{1 - \bar{\alpha}_t}}, \\
\sigma &= \eta \sqrt{\frac{1 - \bar{\alpha}_{t'}}{1 - \bar{\alpha}_t}}
  \sqrt{1 - \frac{\bar{\alpha}_t}{\bar{\alpha}_{t'}}}, \\
x_{t'} &= \sqrt{\bar{\alpha}_{t'}}\, \hat{x}_0
  + \sqrt{1 - \bar{\alpha}_{t'} - \sigma^2}\, \hat{\epsilon}
  + \sigma \mathbf{z},
\label{eq:ddim-step}
\end{align}
where $t' = t_{i+1}$ and $\mathbf{z} \sim \mathcal{N}(0, I)$.  Setting $\eta = 0$
recovers deterministic DDIM; $\eta = 1$ recovers ancestral DDPM sampling.

\paragraph{Rotation step (geodesic interpolation on SO(3)).}
The predicted clean rotation is $\hat{R}_0 = \exp([\mathbf{v}]_\times) \cdot R_t$.
The full rotation from the current state to the predicted clean state is
$\Delta R = \hat{R}_0 \cdot R_t^\top$.  We take a fractional geodesic step:
\begin{equation}
R_{t'} = \exp\!\Bigl(\frac{t_i - t_{i+1}}{t_i} \cdot \mathrm{Log}(\Delta R)\Bigr) \cdot R_t,
\label{eq:geodesic-step}
\end{equation}
where $\mathrm{Log}: \mathrm{SO}(3) \to \mathfrak{so}(3)$ is the matrix
logarithm.  The log map is computed via
$\mathrm{Log}(R) = \frac{\theta}{2\sin\theta}\,\mathrm{vee}(R - R^\top)$
for the general case, with first-order Taylor expansion for $\theta \approx 0$
and an eigenvector-based method for $\theta \approx \pi$ (extracting the
rotation axis from the symmetric part $(R + I)/2$).

\paragraph{Output.}
After $S=100$ steps, translations $t_k^{\mathrm{out}} = x_{t_S}$ and rotations $R_k^{\mathrm{out}} = R_{t_S}$ are output.

\clearpage
\section{Polyformer data for an additional domain 3g0vA00}

\begin{figure}[hbt!]
    \centering
    \includegraphics[width=0.75\linewidth]{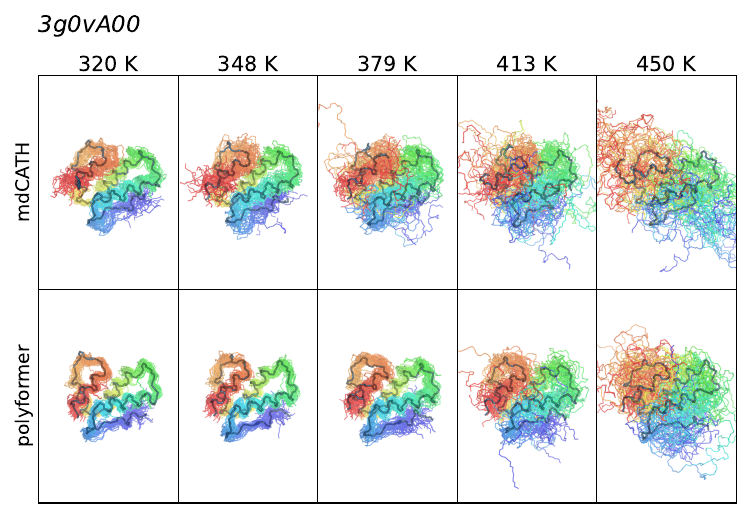}
    \caption{Comparison of Molecular Dynamics and Polyformer sampled conformation ensembles, same as Fig.~\ref{fig:potato} but for domain 3g0vA00.
    }
    \label{fig:potato_3g}
\end{figure}

\begin{figure}[hbt!]
    \centering
    \includegraphics[width=\linewidth]{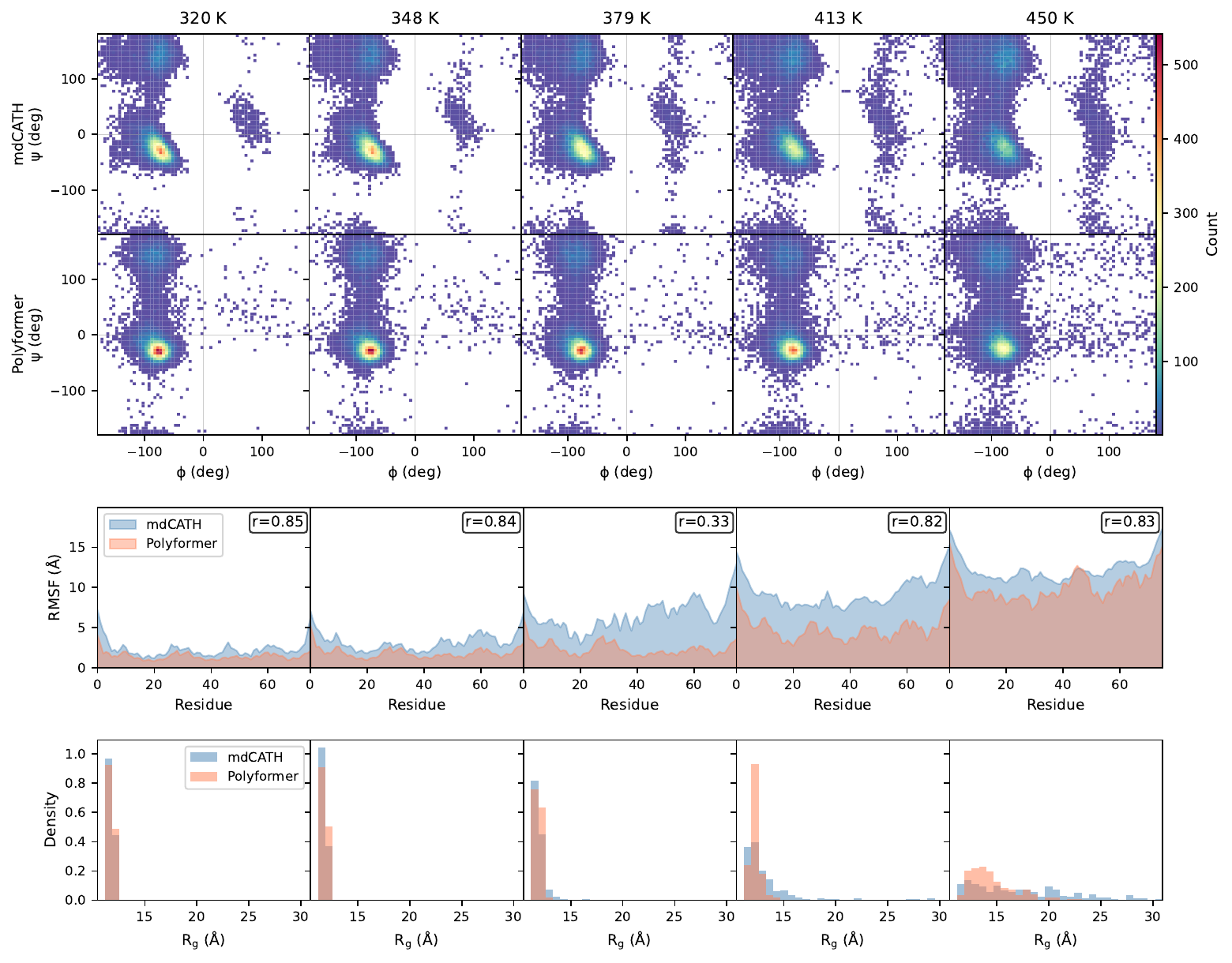}
    \caption{Quantitative comparison of Polyformer and Molecular Dynamics sampled conformation ensemble, same as Fig.~\ref{fig:single_domain_dist} but for domain 3g0vA00.}
    \label{fig:single_domain_dist_3g}
\end{figure}

In Figs.~\ref{fig:potato_3g} and \ref{fig:single_domain_dist_3g} we plot the same data as Fig.~\ref{fig:potato} and \ref{fig:single_domain_dist} but for a different protein domain, 3g0vA00. 

\label{app:more-data}

\clearpage

\section{Polyformer without ESM-2 embedding}
\label{app:no-ESM2}
\begin{figure}[hbt!]
    \centering
    \includegraphics[width=0.75\linewidth]{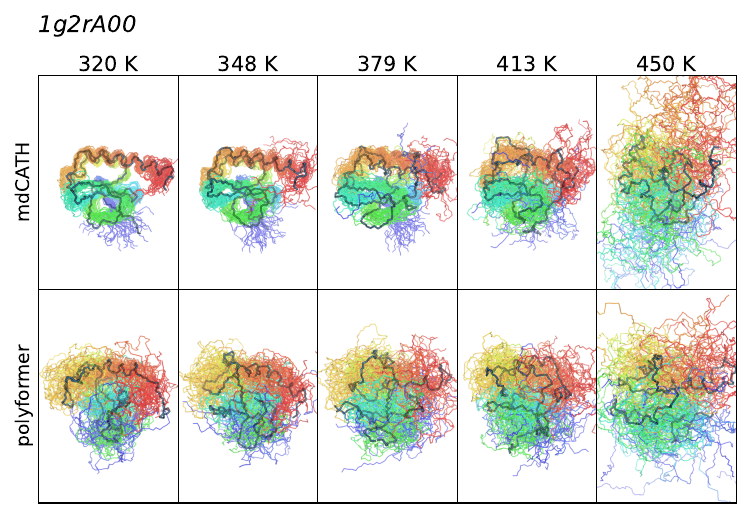}
    \includegraphics[width=0.75\linewidth]{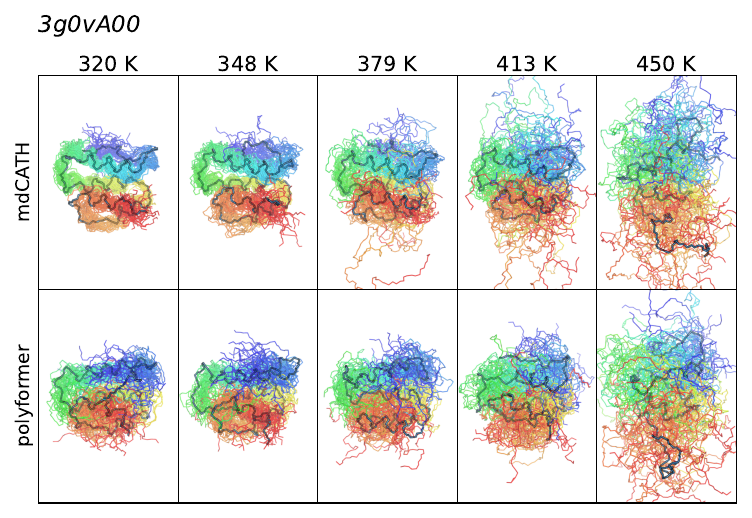}
    \caption{Performance of Polyformer without ESM-2 embedding. Compare with Fig.~\ref{fig:potato} in the main text.
    }
    \label{fig:potato_random}
\end{figure}

\begin{figure}[hbt!]
    \centering
    \includegraphics[width=\linewidth]{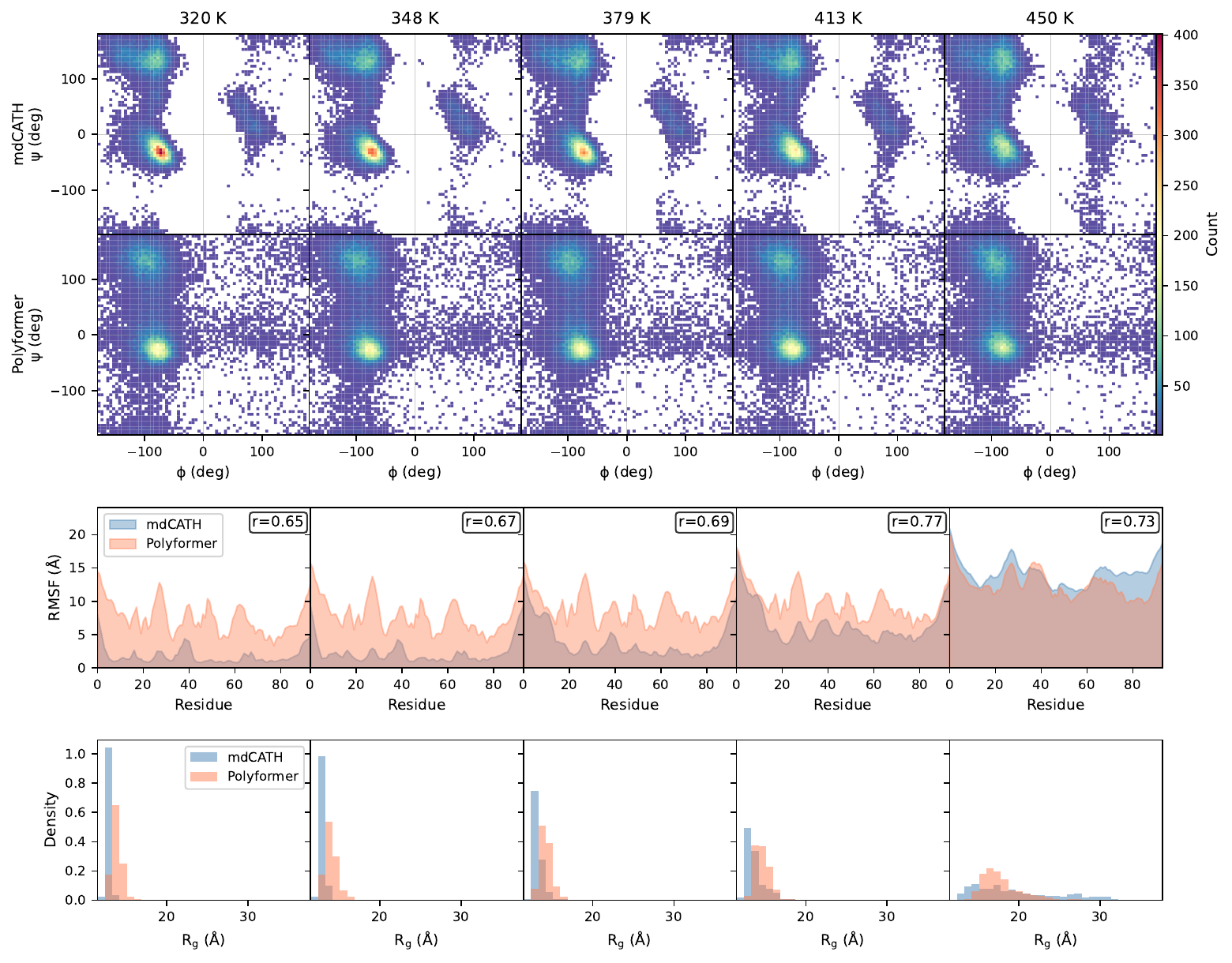}
    \caption{Performance of Polyformer without ESM-2 embedding on domain 1g2rA00. Compare with Fig.~\ref{fig:single_domain_dist} in the main text.}
    \label{fig:single_domain_dist_random}
\end{figure}

\begin{figure}[hbt!]
    \centering
    \includegraphics[width=\linewidth]{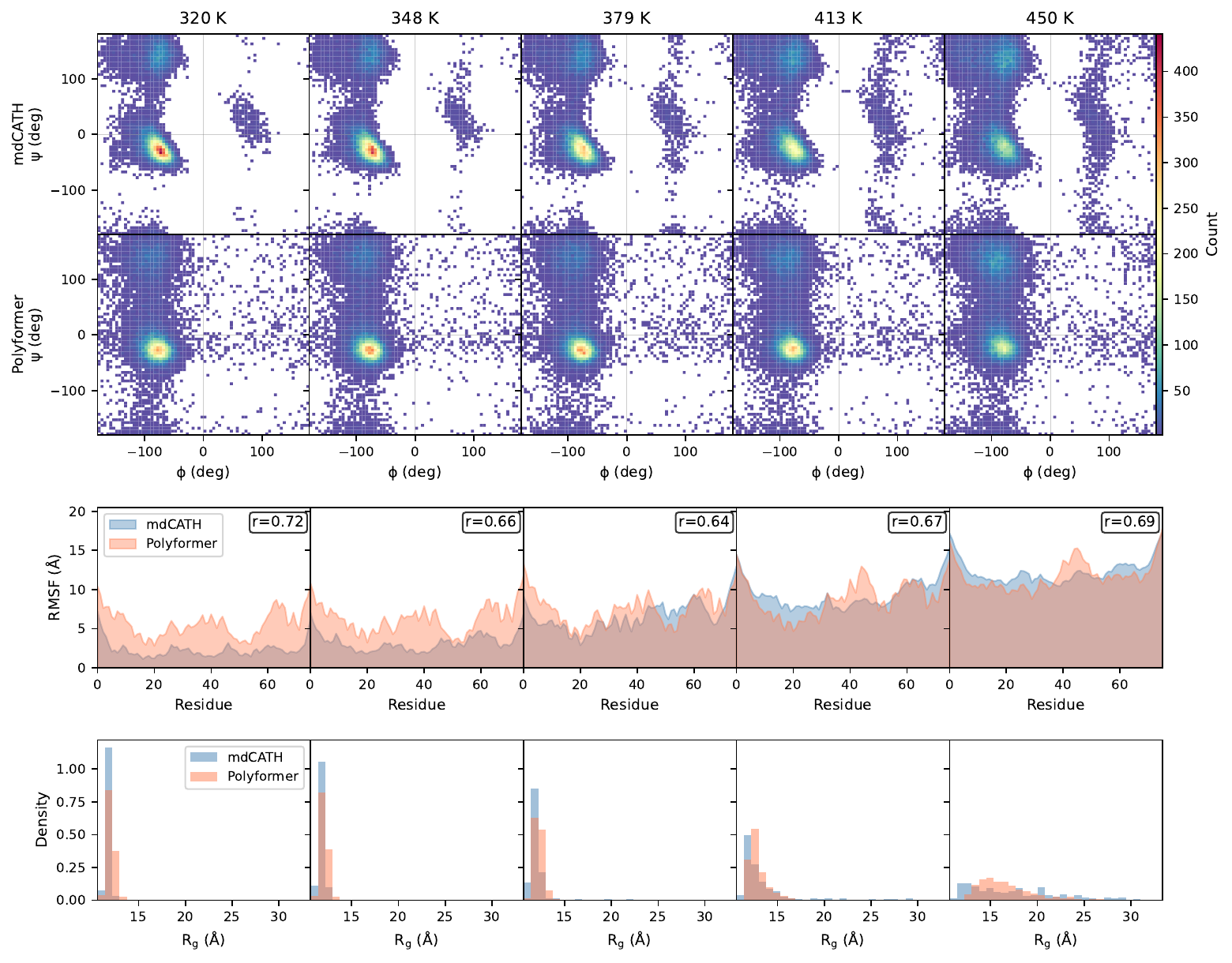}
    \caption{Performance of Polyformer without ESM-2 embedding on domain 3g0vA00. Compare with Fig.~\ref{fig:single_domain_dist_3g} in the appendix.}
    \label{fig:single_domain_dist_random_3g}
\end{figure}

\begin{figure}[hbt!]
    \centering
    \includegraphics[width=\linewidth]{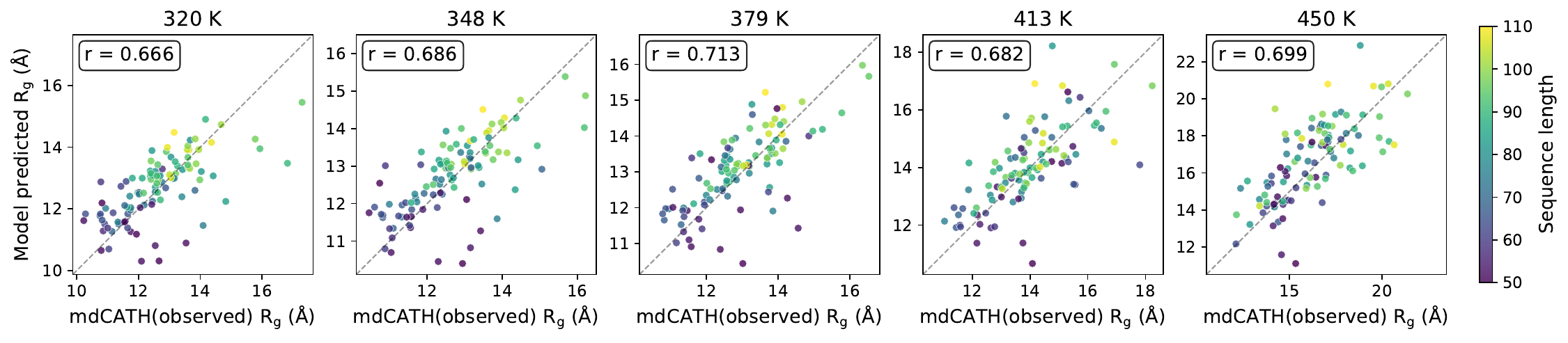}
    \caption{Performance of Polyformer without ESM-2 embedding. Compare with Fig.~\ref{fig:all_domains_rg} in the main text.}
    \label{fig:all_domain_rg_random}
\end{figure}

To understand the role of ESM-2 embedding, we have replaced it with random vector embedding and retrained Polyformer. The performance of Polyformer degrades somewhat, but the model is still able to fold proteins and predict how they denature with temperature. Our results are summarized in Figs.~\ref{fig:potato_random}-\ref{fig:all_domain_rg_random}.

\clearpage

\section{Comparison of Polyformer with SimpleFold}
\label{app:simplefold}
\begin{figure}[hbt!]
    \centering
    \includegraphics[width=0.25\linewidth]{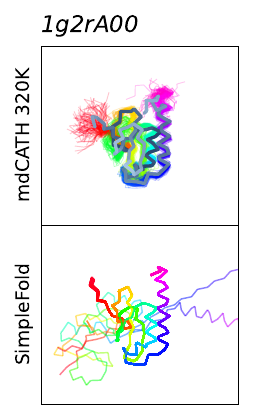}
    \includegraphics[width=0.25\linewidth]{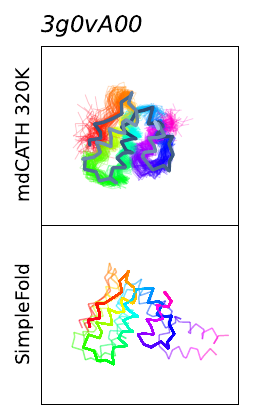}
    \caption{Performance of SimplFold. Compare with Fig.~\ref{fig:potato} in the main text. Because SimpleFold tends to predict one dominant conformation, we have not highlighted the matching pair in the SimpleFold panel. For both domains, SimpleFold predicted two structures that did not match the other 298 strcutures and appear as thinner lines. 
    }
    \label{fig:potato_simplefold}
\end{figure}

Finally, we compare Polyformer to Simplefold. While both are generative models that use DiT architecture, Simplefold was trained on PDB data as opposed to mdCATH data and therefore does not know about thermodynamic ensembles. As a result, SimpleFold tends to predict a single dominant structure for the protein domain, see Fig.~\ref{fig:potato_simplefold}. The dominant SimpleFold structures are quite close to the centroid of mdCATH. On rare occasions SimpleFold does completely misfold the protein domain. We remark that even at low temperatures these domains are expected to be quite flexible and there should not be one dominant conformation.

\clearpage

\begin{figure}[hbt!]
    \centering
    \includegraphics[width=0.3\linewidth]{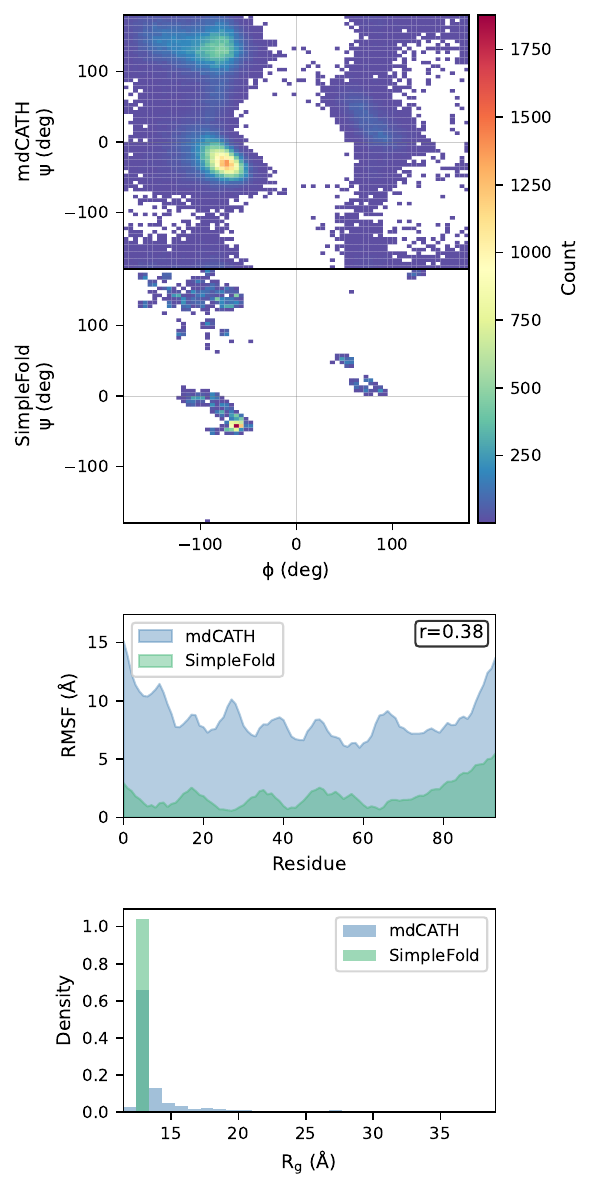}
    \includegraphics[width=0.3\linewidth]{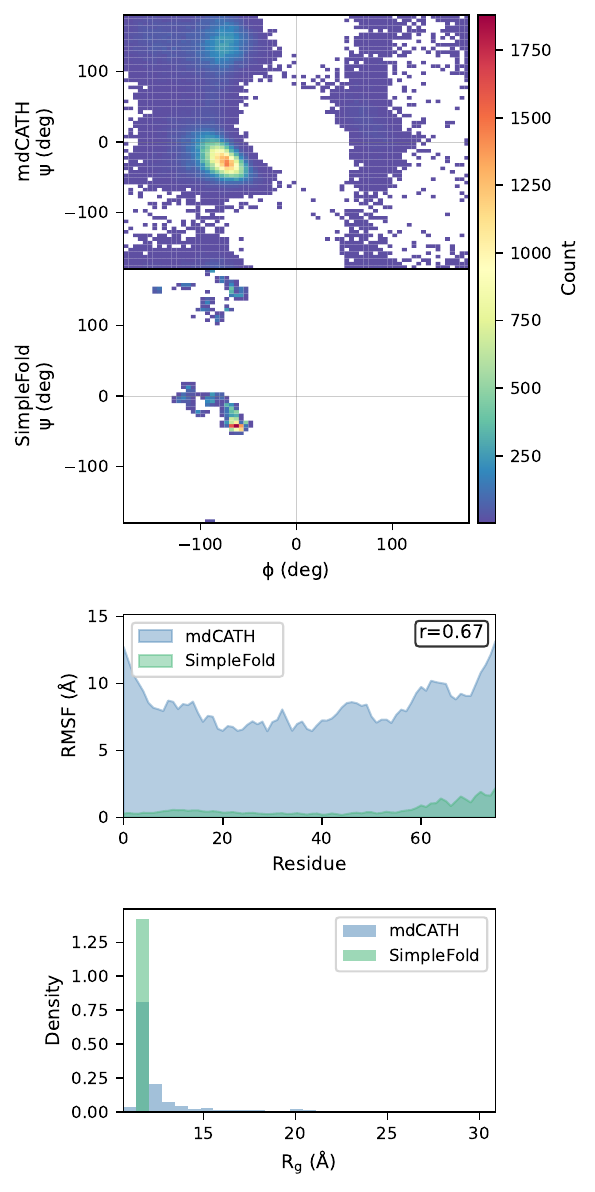}
    \caption{Performance of SimpleFold. Compare with Fig.~\ref{fig:single_domain_dist} in the main text. 
    }
    \label{fig:simplefold}
\end{figure}

SimpleFold predicts a much more rigid structure for proteins as compared to Polyformer. On Ramachandran plot, see Fig.~\ref{fig:simplefold} (top panels), we observe that SimplFold rigidly locks the peptides into alpha-helx or beta-sheet angles. The bimodal distribution of conformations (298 dominat + 2 misfloded) are responsible for the \gl{rmsf} and \gl{rg} plots, see Fig.~\ref{fig:simplefold} (middle and bottom panels)

\clearpage

\section{Reciprocal vectors}
\label{app:fourier}
\begin{figure}[hbt!]
    \centering
    \includegraphics[width=\linewidth]{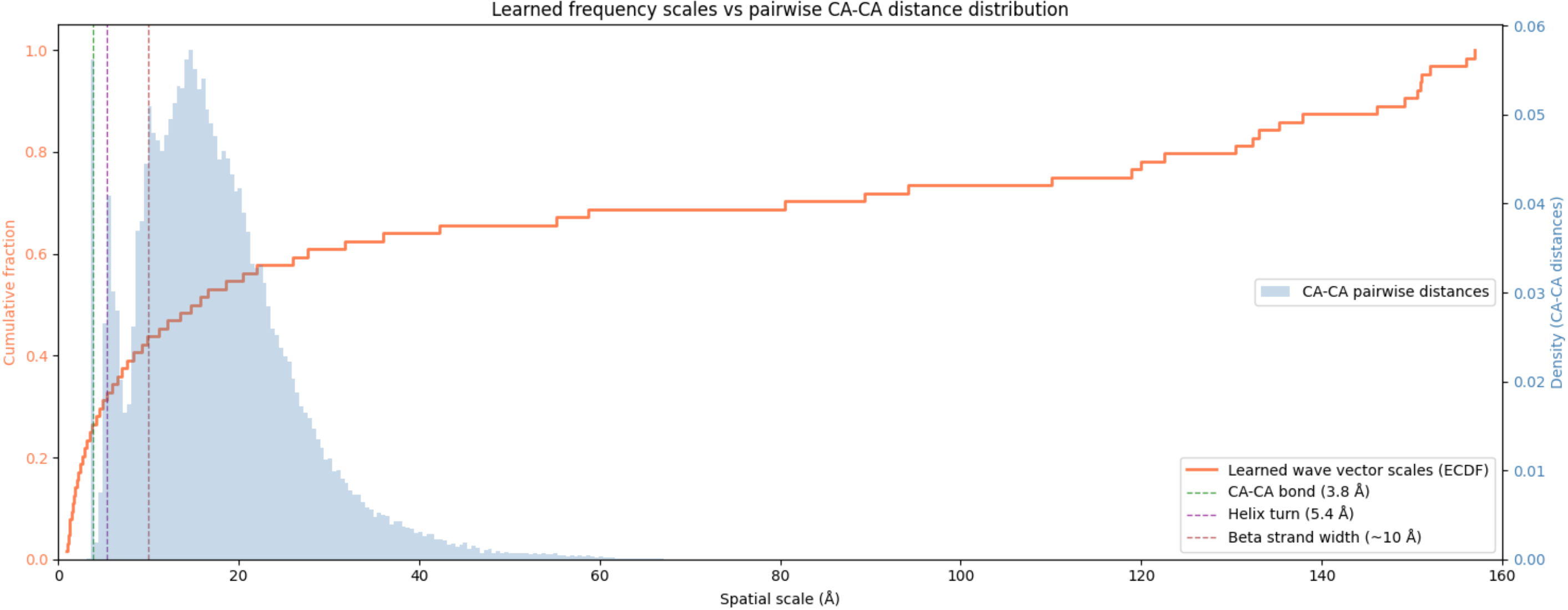}
    \caption{CDF of inverse of the learned reciprocal vectors (orange) and PDF of the C$\alpha$-C$\alpha$ distances shaded blue. Observe that most of the inverse reciprocal vector magnitudes span the range of C$\alpha$-C$\alpha$ distance. }
    \label{fig:fourier}
\end{figure}

We plot the distribution of the learned reciprocal vectors and C$\alpha$-C$\alpha$ distances in Fig.~\ref{fig:fourier}.

\end{document}